\documentclass[aps,prc,showpacs,twocolumn,floatfix,nofootinbib]{revtex4}
\usepackage{epsfig,graphics,color}
\usepackage{graphicx}% Include figure files
\usepackage{dcolumn}% Align table columns on decimal point
\usepackage{bm}% bold math
\usepackage{amsmath}
\usepackage{textcomp}

\newcommand \f {\not\!}

\newcommand \kd  {\delta}
\newcommand \ra  {\rightarrow}

\newcommand \g {\gamma}

\newcommand \e {\epsilon}

\newcommand \x {\cdot}
\newcommand \hf {\frac{1}{2}}
\newcommand \A {\alpha}

\newcommand \lc {\langle}
\newcommand \rc {\rangle}

\newcommand \bvec{\left( \begin{array}{c} }
\newcommand \evec{\end{array} \right)}
\newcommand \tr {\mbox{{ Tr}}}

\newcommand \bea{\begin{eqnarray} }
\newcommand \eea{\end{eqnarray} } 
\newcommand \nn {\nonumber}
\newcommand {\be} {\begin{equation}}
\newcommand {\ee} {\end{equation}}

\newcommand {\mbx} {\mbox{}}

\newcommand {\ata} {& \times &}
\newcommand {\psibar} {\bar{\psi}}
\newcommand {\Q}  {\mathcal{Q}}
\begin{document}

\title{Multiple scattering of heavy-quarks in dense matter and the parametric prominence of drag}
\author{Raktim Abir}
\author{Gagan Deep Kaur}
\author{Abhijit Majumder}
\affiliation{Department of Physics and Astronomy, Wayne State University, Detroit, USA.}

\date{\today}

\begin{abstract} 
The case of heavy quark propagation in dense extended matter is studied in the multiple 
scattering formalism of the higher twist energy loss scheme. We consider the case of 
deep inelastic scattering off a large nucleus.
The hard lepton scatters off a heavy quark fluctuation within one of the nucleons. 
This heavy quark then propagates through the dense medium, multiply scattering off the gluon 
field of the remaining nucleons in its path. We consider the fictitious process where a heavy quark 
propagates through the nucleus without radiation. Invoking Soft-Collinear Effective Theory power counting arguments, 
we consider the case of a ``semi-hard'' heavy-quark where the mass is of the order of the out-going momentum and larger than the transverse momentum 
imparted per unit length due to scattering. 
In this limit, it is found that longitudinal momentum exchanges (quantified by the transport coefficient $\hat{e}$) 
have a comparable effect on the off-shellness of the propagating quark, as the transverse momentum 
exchanges (quantified by $\hat{q}$) which constitute the leading cause of off-shellness for propagating light quarks or gluons. 
Consequences of this new hierarchy for the propagation of the heavy quark in dense matter are discussed.  
\end{abstract}
\pacs {12.38.Mh,12.38.-t,12.38.Cy}

\maketitle 
\section{Introduction}
With the advent of hard sector observables at the LHC \cite{Cole:2011zz,Tonjes:2011zz}, the medium modification of high energy jets has become one of the forefront topics of research. 
The observed suppression in the light flavor sector, especially the dependence on the transverse momentum of the observed hadron, is now well 
understood within a factorized perturbative QCD (pQCD) based approach~\cite{Majumder:2011uk}. 
However, the heavy quark sector has remained a bit of a mystery: There is an observed large 
suppression, both at the Relativistic Heavy-Ion Collider (RHIC) and at the Large Hadron Collider (LHC).

Simple minded extensions of the formalism applied to light quarks
have yielded theoretical results that are in moderate agreement with experimental measurements ~\cite{Zhang:2004qm,Qin:2009gw,Abir:2011jb,Abir:2012pu}. The agreement with measurements of the   
nuclear modification factor and the azimuthal anisotropy of non-photonic leptons (from the decay of open heavy flavor hadrons) at RHIC is found to improve with 
increasing transverse momentum ($p_{T}$) of the detected lepton.
%is considerably large. 
At the LHC, one is not restricted to leptons from the decay of 
$D$ and $B$ hadrons, but instead has access to the nuclear modification factors of both $D$ and $B$ hadrons separately.
The accuracy and $p_{T}$ range of the new measurements of heavy-quark suppression at the LHC, along with the fact that the medium is now both larger and denser calls for a more sophisticated approach. It is the object of this paper to lay the ground work for such an approach based on pQCD, 
in particular, we will incorporate power-counting techniques borrowed from 
Soft-Collinear-Effective-Theory (SCET)~\cite{Bauer:2000yr,Bauer:2001ct,Bauer:2002nz,Bauer:2001yt} to address the issue. 
This paper extends the effort started in 
Refs~\cite{Majumder:2007hx,Majumder:2007ne,Majumder:2008zg,Idilbi:2008vm,Majumder:2009zu,Majumder:2009ge,Qin:2012fua} 
which systematically extends the next-to-leading twist set up of 
Refs.~\cite{Guo:2000nz,Wang:2001ifa} to a scattering resummed formalism for light and (in this paper) to heavy-flavors. 
While gluon radiation from the heavy-quark will not be considered in this paper, the scatterings of the 
heavy quark will engender both longitudinal and transverse momentum transfer, leading to the simultaneous appearance 
of transverse diffusion, and longitudinal drag and straggling~\cite{Majumder:2008zg,Qin:2012fua}. While similar calculations have appeared for a light quark, 
the surprise in this case is the importance of longitudinal transfers codified by $\hat{e}$ to the stimulated off-shellness of the heavy-quark. 
This turned out to be important to the specific case of ``semi-hard'' heavy-quarks: a terminology that will be made clear in the subsequent sections.

To this end, we consider the theoretically well defined case of heavy-quark production and propagation 
in Deep-Inelastic Scattering on a large nucleus (A-DIS), where the produced heavy quark propagates through 
the dense extended and confined nuclear medium. 
The remaining sections are organized as follows: in Sec.~II, we will setup the 
basic formalism of DIS on a large nucleus where the hard virtual photon strikes a heavy-quark created due to high 
$Q^{2}$ fluctuations inside a proton. 
In Sec.~III, we consider the multiple scattering of the produced hard quark and 
introduce power-counting scales similar to SCET. 
In Sec.~IV, we introduce the factorization of the final state scattering from the initial parton distribution function (PDF) 
followed by gradient expansion of the hadronic tensor and its resummation.  
%In Sect.~V we recall the structure functions and discuss specific results 
%which were totally absent in light flavor case. 
We offer concluding discussions and an outlook in Sec.~V.
%\begin{figure}[htbp]
%%\begin{center}
%%  \epsfxsize 80mm
%%\hspace{0cm}
%\resizebox{3in}{2.5in}{\includegraphics{gluon_emit_no_scat_W_mu_nu_lables}} 
%%\vspace{0.25cm}
%    \caption{ The diagram for the hadronic tensor of DIS on a proton with an outgoing quark and a radiated gluon.}
%    \label{fig1}
%%  \end{center}
%\end{figure}

%%%%%%%%%%%%%%%%%%%%%%%%%%%%%%%%%%%%%%%%%%%%%%%%%%%%%%%%%%
%%%%%%%%%%%%%%%%%%%%%%%%%%%%%%%%%%%%%%%%%%%%%%%%%%%%%%%%%%
%%%%%%%%%%%%%%%%%%%%%%%%%%%%%%%%%%%%%%%%%%%%%%%%%%%%%%%%%%

\section{Deep inelastic scattering and intrinsic heavy flavor.}

%%%%%%%%%%%%%%%%%%%%%%%%%%%%%%%%%%%%%%%%%%%%%%%%%%%%%%%%%%
%%%%%%%%%%%%%%%%%%%%%%%%%%%%%%%%%%%%%%%%%%%%%%%%%%%%%%%%%%
%%%%%%%%%%%%%%%%%%%%%%%%%%%%%%%%%%%%%%%%%%%%%%%%%%%%%%%%%%

In this section, we set up the basic formalism and environment where the scattering of the heavy-quark will be derived. 
While the formalism will be set up within the framework of DIS on a nucleus, the scattering of the heavy-quark will be 
factorized both from the initial hard scattering which produces the outgoing heavy-quark as well as from the 
many soft matrix elements which appear in the subsequent multiple scattering. 

Consider the deep-inelastic scattering of a virtual 
photon with a heavy quark off a nucleon, within a large nucleus with mass number $A$. The nucleus possess a momentum $P = pA$, with $p$ the 
average momentum of a nucleon in this nucleus. 
A frame is considered where the exchanged virtual photon has no transverse momentum, and has momentum 
components, 
\bea
q \equiv [q^{+}, q^{-}, q_{\perp}] = \left[  -\frac{-q^2}{2q^{-}} , q^{-} , 0, 0 \right].
\eea 
The nucleons in the nucleus have a mean momentum $p$. We are interested in the reaction, 
\bea
e(L_1) + A(p) \longrightarrow e(L_2) + J_{\Q}(~\vec{l}~) + X ,
\label{chemical_eqn}
\eea
where, $e(L_{1})\,[e(L_{2})]$ represents the incoming (outgoing) electron with momentum $L_{1}\, (L_{2})$, $A(p)$ 
represents the incoming nucleus and $J_{\Q} (~\vec{l}~)$ represents the outgoing jet which contains one heavy quark $\Q$ 
with mass $M$. Since there are no valence heavy-quarks within the nucleon, to produce a jet containing a single heavy-quark, the virtual photon will 
have to strike a heavy quark from within the sea of the nucleon, 
{\it i.e.}, from a $\Q \bar{\Q}$ fluctuation. 
As a result, the outgoing remnants of the nucleon, denoted by the $X$ in the equation above, will contain a $\bar{\mathcal Q}$. 
It is of course equally likely that the $\bar{\mathcal Q}$ will be struck by the virtual photon and the remnants of the proton will contain the 
quark $\mathcal Q$; this will make very little difference to our discussion of multiple scattering of the hard parton as it passes through a nucleus.  
In this article, we will not discuss the dynamics of the production of the heavy quark, containing it within a parton distribution 
function. We high-light the power counting of the momentum components. We consider a quark mass $M \gg \Lambda_{QCD}$ and a final 
outgoing quark momentum which is larger, but of the order of the quark mass. In the frame where the 
proton is boosted by a factor $\gamma = 1/{\lambda}$ in the $`+'$ direction, we have the momentum components of the incoming heavy quark as, 
\bea
p_{\Q} = \left[ p_{\Q}^{+}, p_{\Q}^{-}, \vec{p}_{\Q \perp} \right] \equiv \left[ \sqrt{2} \gamma M, \frac{ M}{2 \gamma \sqrt{2}},  \vec{p}_{\Q \perp}    \right].
\eea
We assume that the quark, anti-quark fluctuation is almost stationary in the proton rest frame and thus $\vec{p}_{\Q \perp} \ra 0 $. The reader should 
note that the boost factor $\gamma$ is simply a alternate variable to $p_{Q}^{+}$ and carries no extra information other than the relation between a large 
$p_{\Q}^{+}$ and a small $p_{\Q}^{-}$.

The momentum components of the incoming photon are assumed to be,
\bea
q = \left[ - \sqrt{2} \gamma M   + \frac{M^{2}}{2q^{-}} ,  q^{-} - \frac{ M}{2\gamma\sqrt{2}}, 0   \right] .
\eea
In the equation above, we are assuming that $\gamma M \gg M \sim q^{-} \gg  M/\gamma $. Thus we have $Q^{2} \simeq \sqrt{2}\gamma M q^{-}   $. 
As a result, we obtain the final out-going quark to have momentum components 
\bea
(p_{\Q} + q) \simeq \left[   \frac{M^{2}}{2q^{-}} , q^{-} , 0 \right].
\eea
For concreteness, we may consider $M \lesssim q^{-} \sim \sqrt{\lambda} Q$ for slow heavy quarks. 

Given these simplifications, we may express the differential cross section to produce a hard parton with 3-momentum $\vec{l} \equiv l^{-}, l_{\perp}$ 
in the DIS on a large nucleus as, 
\bea
\frac{E_{L_2}d\sigma} {d^3 L_2 dl^{-} d^2l_\perp} &=&
\frac{\A_{e}}{2\pi s}~\frac{1}{Q^4} ~L_{\mu \nu}~\frac{d W^{\mu \nu}}{d l^{-} d^2 l_\perp}, \label{LO_cross}
\eea
where, the Mandelstam variable, $s = (p+L_1)^2$. All terms that contain the wave-functions of the incoming and outgoing 
leptons are included in the leptonic tensor,
\bea 
L_{\mu \nu} = \frac{1}{2} \tr [ \f L_1 \g_{\mu} \f L_2 \g_{\nu}].
\eea
The initial state of the incoming large nucleus, with $A$ nucleons and an average momentum $p$ per nucleon, is represented by the ket $| A; p \rc$. The 
final unidentified hadronic or partonic state is defined  as 
$| X \rc $. The entire strongly interacting part of the cross section is included in the hadronic tensor, 
defined as 
\bea 
&& W^{\mu \nu} = \sum_X (2\pi)^{4} \kd^4 (q\!+\!P_A\!-\!p_X ) \nn \\
&& \times  \lc A; p |  J^{\mu}(0) | X  \rc \lc X  | J^{\nu}(0) | A;p \rc  \nn \\
&&= 2 \mbox{Im} \left[  \int d^4 y e^{i q \cdot y } \lc A;p | J^{\mu} (y) J^{\nu}(0) | A;p \rc \right]
\eea
where the sum ($\sum_X$)  runs over all possible hadronic states and $J^{\mu}$ is the 
hadronic current ($J^{\mu} =  Q_\Q \bar{\psi}_\Q \g^\mu \psi_\Q$, where $Q_\Q$ is the 
charge of the heavy-quark of flavor $\Q$ in units of the electron charge $e$). 
Factors of the electromagnetic coupling constant have already been extracted and 
included in Eq.~\eqref{LO_cross}.

Ignoring all power corrections of the order of  $\Lambda_{QCD}/Q$ as well as factors of the heavy quark mass, the hadronic tensor may be expressed 
as (we also take the average over initial states and 
sum over final states to obtain)
%
%\bea
%
%{W_0^A}^{\mu \nu} 
% &\equiv & 2 Im \left[ \int d^4y e^{iq \x y} \lnuc  J^{\mu}(y) J^\nu (0)  \rnuc \right] \nn \\
%
%&=&  C_p^A W_0^{\mu \nu} \label{w_mu_nu_twist=2}\\
%
%\tr \left[ \lnuc \psi(0) \psibar(y) \rnuc  \g^{\mu} (\f p + \f q) 2 \pi \kd [(p+q)^2]  \rnuc \right]
%
%&=& C_p^A  \frac{2 \pi x_B}{2 Q^2} \left[ \tr \left\{ \f{p} \g^\mu \left( \f{q} + x_B \f{p} \right) \g^\nu  \right\}  \right] \nn \\
%\ata \sum_\Q Q_\Q^2f_q (x_B)  \nn \\
%
%&= &C_p^A 2 \pi \left[ g^{\mu -} g^{\nu +} + g^{\mu +} g^{\nu -} - g^{\mu \nu} \right]  \sum_q Q_\Q^2 \nn \\
%
%\ata   \int \frac{d y^-}{2\pi} e^{-ix_B p^+ y^-} \hf \lc p| \psibar(y^-) \g^+ \psi(0) | p \rc \nn.
%
%\eea
%In the equation above, any mass dependent corrections in the hard scattering cross section have been ignored. Our intention is to 
%factorize the hard cross section and parton distribution function from the final scattering of the heavy quark, and then only study 
%the final state propagation of the heavy-quark. In this effort, we only consider the propagation of a heavy-quark without 
%radiation; while future efforts will consider gluon radiation form the heavy quark, we will continue to factorize and ignore mass corrections 
%to the hard cross section. 

\bea
{W_{0}^A}^{\mu \nu} \!\!\!\!\!\!\!&=& C_p^A~ W_0^{\mu \nu} \label{w_mu_nu_twist=2_light} 
=   C_p^A \frac{2 \pi }{2 Q^2} \sum_{\Q}~Q_{\Q}^2~f_{\Q} (x_B) \\
\ata\tr \left[  \f{p}_{\Q} \g^\mu\! \left(\!\f{p}_{\Q}+ \f{q}    \right)\! \g^\nu  \right]    \nn \\
&=&  \sum_{\Q} Q_{\Q}^2 C_p^A~ 2 \pi \left[ g^{\mu -} g^{\nu +} + g^{\mu +} g^{\nu -} - g^{\mu \nu} \right] \nn \\
\ata \int \frac{d y^-}{2\pi} e^{-ix_B p^+ y^-} \hf \lc p| \psibar(y^-) \g^+ \psi(0) | p \rc \nn.
\eea
In the absence of quark mass corrections, the only surviving components of the hadronic tensor are 
${W_0^A}^{\perp \perp}$. However the situation is different for the case of heavy quarks.  Here the components of the hadronic tensor can be expressed as follows,
%\begin{widetext}
\bea
&& {{W_{0}^A}^{\mu \nu}}_{\!\!\!\!\!\!\!\!\!\!M}  \nn \\
&=& C_p^A~ {W_0^{\mu \nu}}_{\!\!\!\!\!\!M} \label{w_mu_nu_twist=2}\\
&\simeq& C_p^A \frac{2 \pi }{2 Q^2} \sum_{\Q}~Q_{\cal Q}^2~f_{\cal Q} (x_B)  \nn \\ 
\ata \tr \left[  (\f{p}_{\Q})~\g^\mu~\left(\f{p}_{\Q}+\f{q}\right)~\g^\nu  \right]  \nn \\
%%
%&=& \sum_{\cal Q} Q_{\cal Q}^2C_p^A~ \frac{2 \pi }{2 Q^2} \nn \\
%%
%\ata \left[ 8p_0^\mu p_0^\nu + 4\left(p_0^\mu q^\nu+ p_0^\nu q^\mu\right) - 4 p_0 \cdot q    g^{\mu\nu}\right] \nn \\
%%
%\ata \int \frac{d y^-}{2\pi} e^{-ix_B p^+ y^-} \hf \lc p| \psibar(y^-) \g^+ \psi(0) | p \rc . \nn \\
%
&\simeq& C_p^A \frac{2 \pi \sqrt{2}\gamma M }{2 Q^2}  \left[  -4 q^- 
%-4p_0^+q^-+2M^2\right)
g^{\mu}_{~\perp} g^{\nu}_{~\perp}
+ \frac{ 4M^2}{q^-} 
%+\frac{4 M^2 p^+}{q^-}
g^{\mu}_{~+}g^{\nu}_{~+}  \right]  \nn \\
%
%&& ~~~\left.  + \left(4\sqrt{2}\gamma M p_0^{\perp}\right)(g^{\mu-}g^{\nu \perp})+(4M^2)(g^{\mu-}g^{\nu+}) + {\cal O}(\lambda)\right]\sum_{\cal Q} Q_{\cal Q}^2 
%
\ata \sum_{\Q} Q_{\Q}^2 \int \frac{d y^-}{2\pi} e^{-ix_B p^+ y^-} \hf \lc p| \psibar(y^-) \g^+ \psi(0) | p \rc . \nn
\eea
In the equation above, all terms suppressed by powers of $\lambda$ have been neglected. 
We have two leading components of the hadronic tensor $W^{\perp \perp}$ and $W^{++}$. 
In what follows, we will study the modification of these terms in an extended nuclear medium.

In the above discussion, we have introduced the dimensionless small parameter $\lambda$. 
This is a concept borrowed from soft collinear effective theory (SCET)~\cite{Bauer:2000yr,Bauer:2001ct,Bauer:2002nz,Bauer:2001yt}, and constitutes the power counting variable: Terms that are sub-leading in $\lambda$ will be dropped. A similar program 
for light quarks was carried out in Ref.~\cite{Idilbi:2008vm}. 
As mentioned above, in the remainder of this paper we will consider the case of a heavy quark with a momentum 
$p \sim \sqrt{\lambda}Q$ and mass $M \sim \sqrt{\lambda}Q$. 
The off-shellness of the hard virtual photon $Q$ is the hardest scale in the problem and hard momentum components are 
expected to scale as $\sqrt{\lambda}Q$. Softer momentum components may scale as $\lambda Q$ or even $\lambda^{2} Q$.
The meaning of $\sqrt{\lambda}$ is an intermediate suppression factor that may not be neglected. 
Note that at this level of approximation, there is no shift in the Bjorken variable $x_{B} $, given as 
\bea
x_{B} = \frac{Q^{2} }{2 p^{+} q^{-}}  \sim \frac{Q^{2}}{ [Q / \sqrt{\lambda}] \sqrt{\lambda} Q  } . \label{x-B}
\eea
Thus $\mathcal{O} (x_{B}) \sim 1 $, {\it i.e.}, $x_{B}$ is a large momentum ratio. 
%This is caused by the fact that the incoming quark has a negative light-cone momentum $p^{-} = M^{2} / (2 p^{+})$, 
%i.e., unlike a light quark, the heavy quark will have a non-vanishing $p_{0}^{-}$. 

In the subsequent sections, we will consider both  the hadronic tensor components, ${W_0^A}^{\perp \perp}$ and ${W_0^A}^{+ +}$ and investigate how they evolve in the final state propagation of the heavy quark. Radiation will be ignored. In all cases, the initial state parton distribution function and 
hard cross section will be defined as above. The focus will be only on the final state 3-momentum distribution of the heavy quark, which,
in the case without any final state scattering will possess a narrow width and can be approximated as a delta function, {\it i.e.},
\bea
\frac{d {W_{0}^A}^{\mu \nu}}{dl^{-} d^2l_\perp} && = {W_{0}^A}^{\mu \nu} {\phi_{0}} (l^{-}, l_\perp)  \nn \\
&& = {W_{0}^A}^{\mu \nu} \kd(l^{-} - q^{-} ) \kd^{2} (l_{\perp})~.
\eea
In the subsequent section, we will consider only the modification to the final state momentum distribution $\phi (l^{-}, l_\perp)$.

%%%%%%%%%%%%%%%%%%%%%%%%%%%%%%%%%%%%%%%%%%%%%%%%%%%%%%%%%%
%%%%%%%%%%%%%%%%%%%%%%%%%%%%%%%%%%%%%%%%%%%%%%%%%%%%%%%%%%
%%%%%%%%%%%%%%%%%%%%%%%%%%%%%%%%%%%%%%%%%%%%%%%%%%%%%%%%%%

\section{Multiple Scattering and Final state Power Counting.}

%%%%%%%%%%%%%%%%%%%%%%%%%%%%%%%%%%%%%%%%%%%%%%%%%%%%%%%%%%
%%%%%%%%%%%%%%%%%%%%%%%%%%%%%%%%%%%%%%%%%%%%%%%%%%%%%%%%%%
%%%%%%%%%%%%%%%%%%%%%%%%%%%%%%%%%%%%%%%%%%%%%%%%%%%%%%%%%%

In the preceding section, the final state momentum distribution of the outgoing quark has been identified and factorized from 
the hard cross section. In this section, the propagation of this quark without radiation will be considered. This is by no means 
a physical process. High momentum partons produced in hard processes are most often produced far off their mass shell {\it i.e.}, with a considerable virtuality $\mu^{2}$, which, though small compared to the forward energy of the quark, is still much larger than $\Lambda_{QCD}^{2}$. The hard parton tends to shed this large virtuality through a series of gluon emissions. The emitted gluons are also virtual and will radiate further leading to the development of a partonic cascade. Each parton in this cascade will engender multiple scattering while radiating a gluon (or splitting into a quark antiquark pair). In order to incorporate the effect of multiple scattering on the development of a cascade, the propagation of single partons in a medium, without emission, has to be clearly understood. In prior efforts, in Refs.~\cite{Majumder:2007hx,Majumder:2008zg,Idilbi:2008vm,Qin:2012fua}, the effect of multiple scattering on a light parton has 
been derived. In order to calculate the development of a cascade from a heavy quark one has to calculate the effect of multiple scattering on a single heavy quark that does not radiate. This will be carried out below.

As the struck heavy quark propagates through the nucleus, it will scatter off the dense gluon field within nucleons in its path. 
Every scattering engenders an extra factor of the strong coupling constant $\alpha_{s}$. As will be shown later, every scattering also contains an integration over the location of where the scattering took place. In a large nucleus $(A \gg 1)$, these length integrals give large factors that counter the suppression due to the appearance of factors of $\A_{s}$.

\begin{widetext}

\begin{figure}[htbp]
\begin{center}
  \epsfxsize 80mm
\hspace{0cm}
\resizebox{5.2in}{2in}{\includegraphics{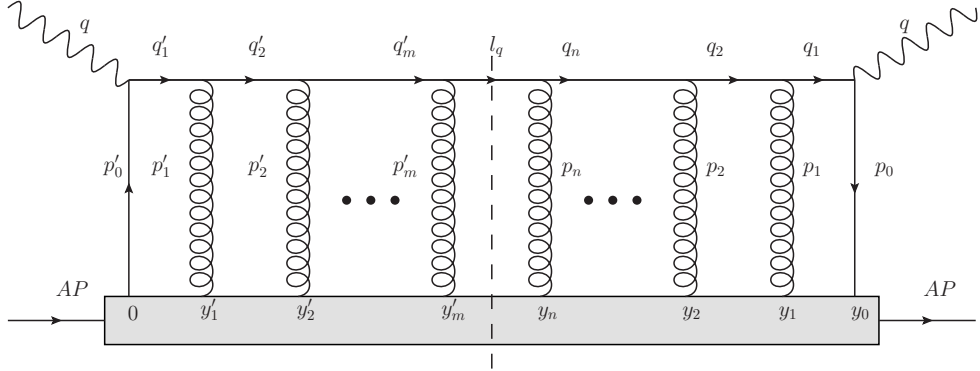}} 
\vspace{0.25cm}
\caption{A heavy-quark produced in DIS on a nucleon inside a large nucleus. The produced quark then propagates through the remaining nucleus multiply scattering off the soft gluon field of the nucleons behind the struck nucleon.}
    \label{mult-scat}
  \end{center}
\end{figure}

\end{widetext}

%\begin{figure}[htbp]
%\begin{center}
%  \epsfxsize 80mm
%\hspace{0cm}
%\resizebox{3.2in}{2in}{\includegraphics{dis_n_label}} 
%\vspace{0.25cm}
%\caption{ The leading length enhanced contribution for $2n$ scatterings on the quark. }
%    \label{mult-scat}
%  \end{center}
%\end{figure}

In order to remain consistent with previous calculations, we will denote $p_{\Q}$ as $p_{0}$ in the remainder of this paper. Thus the incoming heavy quark possesses a momentum, $p_{0}$ such that, 
\bea
p_{0} &\equiv& \left[  {p}_{0}^{+} , \frac{M^{2}}{2{p}_{0}^{+}} + \delta p_{0}^{-}, 0, 0 \right]  \\
& =&  \left[ x_{B} p^{+}  , \frac{M^{2}}{ 2 x_{B}p^{+}} + \delta p_{0}^{-} ,0,0\right]. \nn
\eea
In the equation above, $p_{0}^{-} = M^{2}/(2p_{0}^{+}) + \delta p_{0}^{-}$, where $M^{2}/2p_{0}^{+} \sim \lambda^{3/2} Q$ and 
$\delta p_{0}^{-} \sim \lambda^{2} Q$ is considered as a small correction. 
 The incoming virtual photon has momentum components 
\bea
q \simeq \left[ -\frac{Q^2}{2q^-} + \frac{M^{2}}{2q^{-}} , q^-, 0,0 \right],
\eea
with $Q^2 \simeq 2\gamma Mq^-$ and thus the final outgoing heavy quark has momentum components, 
\bea
q_1&=&p_{0}+q \nn \\
&\simeq& \left[x_{B} p^{+}  -\frac{Q^2}{2q^-} + \frac{M^{2}}{2q^{-}} , q^- , 0, 0 \right].
\eea 
In the equation above, we have ignore the small corrections from $p_{0}^{-}$. 
%Current conservation conditions $q^\mu W_{\mu\nu}=0$ and $q^\nu W_{\mu\nu}=0$ require this outgoing 
%heavy quark must strictly be on shell. 
Note that, the vanishing transverse components of the virtual photon are a choice of coordinate system, whereas the 
vanishing components of the incoming quark are an approximation. Insisting that the outgoing quark is 
close to mass shell, this yields $x_{B}$ (identical to $x_{0}$, defined later) as given in Eq.~\eqref{x-B}.

The relevant Feynman diagram is shown in Fig.~\ref{mult-scat}. This describes the hard scattering processes when a virtual photon strikes a heavy quark off the nucleus with momentum $p_0'$ ($p_0$ in amplitude) at space-time point $y_0'=0$ ($y_0$ in complex conjugate amplitude).
It is then scattered back through the nucleus with momentum $q_1'$ ($q_1$ in the complex conjugate amplitude).
While propagating through the large nucleus, 
the heavy quark scatters with the the gluon fields inside the nuclear medium at space-time points $y_j'$ with $0<j<m$ ($y_i$ for the complex conjugate amplitude with $0<i<n$). 
All calculations will be carried out in $A^{-} = 0$ gauge. The choice of gauge controls the power counting of the gluon vector potentials.
As been stated earlier we are only considering the case where the heavy quark traverses the
medium without any radiation. Heavy quark propagation with bremsstrahlung radiation will addressed in future.
In each scattering, the ``semi hard'' heavy quarks accumulates momenta $p_j'$ ($p_i$ in the complex conjugate).
Momentum conservation, at each vertex, enable one to assign various momenta as follows (See Fig.~\ref{mult-scat}):
\begin{eqnarray}
&& q_{i+1} = q_i + p_i = q + \sum_{j=0}^i p_j = q + k_i, \nonumber\\
&& q'_{i+1} = q'_i + p'_i = q + \sum_{j=0}^i p'_j = q + k'_i,
\end{eqnarray}
where we have defined new momentum variables, for convenience, as $k_i = \sum_{j=0}^{i} {p_j}$, and $k_i' = \sum_{j=0}^{i} {p_j'}$, which denote the `total' momentum exchanged between the propagating heavy quark and the nuclear medium. The hadronic tensor now can be written as,

\begin{widetext}
\begin{eqnarray}
 W_{mn}^{A\mu\nu} 
\!\!&=&\!\! \sum_{\Q} Q_\Q^2 g^{n+m} \frac{1}{N_c} {\rm Tr}\left[\left(\prod_{i=1}^{n} T^{a_i}\right) \left(\prod_{j=m}^{1} T^{a'_j}\right) \right]
\int \frac{d^4l_q}{(2\pi)^4} (2\pi)\delta^{+}(l_q^2-M^2)\int d^4y_0 e^{iq\cdot y_0} 
\nonumber \\
&\times&   \left(\prod_{i=1}^{n} \int d^4y_i\right) \left(\prod_{j=1}^{m} \int d^4y'_j \right)\left(\prod_{i=1}^{n} \int \frac{d^4q_i}{(2\pi)^4} e^{-iq_i\cdot (y_{i-1} - y_i)} \right) e^{-il_{q}\cdot (y_{n} - y'_{m})}  \left(\prod_{j=1}^{m}  \int \frac{d^4q'_j}{(2\pi)^4} e^{-
iq'_j\cdot (y'_j - y'_{j-1})}\right)
\nonumber \\
&\times&   \langle A | \bar{\psi}(y_0) \gamma^\mu \left(\prod_{i=1}^{n} \frac{\gamma\cdot q_i+M}{q_i^2 -M^2 - i\epsilon} \gamma\cdot A^{a_i}(y_i)\right) \gamma\cdot l_q
\left(\prod_{j=m}^{1} \gamma\cdot A^{a_j'}(y_j') \frac{\gamma\cdot q'_j+M}{{q'_j}^2 -M^2 + i \epsilon}\right) \gamma^\nu \psi(0) |A\rangle.
\label{main_eq_1}
\end{eqnarray}

\end{widetext}
Here, $Q_{\Q}$, $N_{c}$ and $T^{a_{i}} (T^{a'_{j}})$ are the electromagnetic charge, number of colors, and the SU(3) generators respectively. To make expression simple, we will now change the integral variables $q_{i+1} \to p_i$ and $q_{j+1}' \to p_j'$, and will incorporate one additional exchanged momentum $p_n$ inside the complex conjugate amplitude by bringing the following $\delta$-function,
\begin{eqnarray}
1 = \int \frac{d^4p_n}{(2\pi)^4} (2\pi)^4 \delta^4(l_q - q - p_n).
\end{eqnarray}
%

%\begin{figure}[thb]
%\includegraphics[width=0.95\linewidth]{ht_ms.eps}
% \caption{An order of $n+m$ contribution to hadronic tensor $W^{\mu\nu}$ with $n$ gluon insertions in the complex conjugate and $m$ gluon insertions in the amplitude.
%} \label{mntwist}
%\end{figure}

%At  this point it is important to outline the power counting scales which carries the various approximations that will be carried out in next. The transverse components of exchanged momenta, between the heavy quark and the medium,  scale as $p_{i}^{\perp} \sim \lambda Q$. 
%Due to the large Lorentz boost, the $(+)$-component of $p_{i}$ though quite large, else, due to the requirement that the propagating parton not go off-shell by more than $\lambda Q$, one obtain $p_{i}^{+} \sim \lambda^{2} Q$ for both light and heavy quarks. 
%We will also keep insisting that $p_{i}^{-} \sim \lambda^{2} Q $. One should note that $p_{i}^{-} \sim \lambda Q$ is a perturbatively hard radiation, though soft compared to the jet. 

Mass modifications to the case of a light quark occur from two sources: the pole structure of the propagators, as specified by the denominators and the spin structure from the numerators of the various propagators. The sum over spins in the numerator of each quark propagator has a factor of $M$. This is an obvious $M$ dependent correction compared to the case of mass-less quarks. However, to include factors of $M$ require that there be at least 2 factors of $M$  in the trace (else we will have an odd number of $\g$ matrices) and, secondly, each factor be either preceded or followed by a $\g_{\perp} \cdot A_{\perp}$, as 
\bea
\tr \left[   \ldots \g^{-} A^{+}  M \g^{-} A^{+} \!\!\!\ldots \g^{-} A^{+} M \g^{-} A^{+}\!\!\! \ldots \right] = 0.
\eea
This is due to the fact that in light-cone coordinates $\{ \g^{+},\g^{+} \} = \{ \g^{-} , \g^{-} \} = 0 $. As a result, the first non-vanishing correction, from $M$ dependent terms in the numerator, yields an additive contribution to the 
heavy-quark hadronic tensor, of the following form: 
\bea
\mbx\!\!\!\kd W_{\Q}^{\mu \nu} \!\! \propto \!\!\!\tr \left[   \ldots \g_{\perp} A_{\perp}  M \ldots \g_{\perp} A_{\perp} M  \ldots \right]\! \sim\! \lambda^{2} W_{\Q}^{\mu \nu}.
\eea
In the equation above $W_{\Q}^{\mu \nu}$ is the leading contribution to the heavy-quark hadronic tensor. The overall factor of $\lambda^{2}$ is due to the appearance of two factors of $A_{\perp}$ which scale as $\lambda A^{+}$, in $A^{-} =0$ gauge~\cite{Idilbi:2008vm,Majumder:2009ge}. 

We now turn to the leading contribution: $W_{\Q}^{\mu \nu}$. To simplify the denominator, consider the denominator of the first propagator in Eq.~\eqref{main_eq_1}, where $q_{1} = q + p_{0}$. This can be expressed as 
\bea
q_{1}^{2} - M^{2} &=& -Q^{2} + 2 q^{+} p_{0}^{-} + 2 q^{-} p_{0}^{+} . 
\eea 
Contour integration on $p_{0}^{+}$ will set $p_{0}^{+} = ( Q^{2}  - 2 q^{+} p_{0}^{-})/(2 q^{-})$. Even if the incoming quark were on mass shell, $p_{0}^{-} = M^{2}/(2 p_{0}^{+})$, a term of order $\lambda^{3/2}Q$ and thus negligible compared to $q^{-} \sim \lambda^{1/2} Q$.

The fate of the remaining denominators, on contour integration, can now be easily surmised. For example, the second propagator yields the relation,
\bea
q_{2}^{2} - M^{2} &= & (q_{1} + p_{1})^{2}  - M^{2} \nn \\
&=& 2p_{1}^{+} p_{1}^{-}  - {p_{1}}_{\perp}^{2} + 2 p_{1}^{+} (q^{-} + p_{0}^{-})  \nn \\
&+& 2  p_{1}^{-} (q^{+} + p_{0}^{+}).
\eea
As in previous calculations with light quarks in Refs.~\cite{Majumder:2007hx,Majumder:2008zg,Qin:2012fua}, we will assume that all rescattering of the produced quark with the soft gluons off the medium engender momentum exchanges where $p_{\perp} \sim \lambda Q$ and $p^{-} \sim \lambda^{2} Q$, with $p^{+}$ is fixed by the requirement that the propagating quark be close to its mass shell. Neglecting all but the lowest power of $\lambda$, we obtain, 
\bea
p_{1}^{+} = p_{1}^{-} \frac{ q^{+} + p_{0}^{+} }{ q^{-}  + p_{0}^{-} }  - \frac{ {p_1}_{\perp}^{2} }{ q^{-}  + p_{0}^{-} }. \label{p-1+}
\eea
Comparing with the results for $p_{i}^{+}$ in Refs.~\cite{Majumder:2007hx,Majumder:2008zg,Qin:2012fua}, the above equation represents a remarkable departure from the case of a light quark. 
In this case of the heavy quark, if $p_{i}^{-} \sim \lambda^{2} Q$, $p_{i}^{+}$ is comparably controlled by $p_{i}^{-}$ and ${p_{i}}_{\perp}$. If $p_{i}^{-} \sim \lambda^{2} Q$ then both terms on the right hand side of Eq.~\eqref{p-1+} are 
present [note that we will have similar expressions for all $i$ as in Eq.~\eqref{p-1+}]. Thus $p_{i}^{+} \sim \lambda^{3/2} Q$, and depends, non-negligibly, on the value of the longitudinal exchange $p_{i}^{-}$. In the case of a light quark, $p_{i}^{+}$ was dominantly controlled by ${p_{i}}_{\perp}^{2}/(2q^{-})$ (with sub-leading corrections from $p_{i}^{-}$) and in that case $p_{i}^{+} \sim \lambda^{2}Q$.

As a result of the above considerations, the $(+)$-components of all the exchanged gluons off which the heavy-quark scatters can be of the order of $\lambda^{2} Q$ or even as high as $\lambda^{3/2} Q$. This implies that if the heavy-quark goes off shell and radiates a perturbatively resolvable radiation, it can go off shell by $\delta q \sim \lambda^{2} Q$. 

Since the two light-cone components of the heavy-quark momentum are of the order $\sqrt{\lambda} Q$, the radiation pattern (and the energy loss) from a heavy-quark driven off shell by scattering may be somewhat different than for a light quark and gluon. The radiation from a heavy-quark undergoing multiple scattering will be discussed in a future effort. In the current effort we will lay the ground work for this calculation by elucidating the propagation of a single heavy-quark in an extended medium without emission. 

In the subsequent section, Eq.~\eqref{main_eq_1} will be expanded in a power series in $\lambda$, where we will find that 
the leading terms in the series expansion will yield both a transverse diffusion equation and a longitudinal drag equation, 
and not just a diffusion equation as in the case of a light quark~\cite{Majumder:2007hx}.
 
We now study the structure of the denominator of an arbitary propagators. For the heavy-quark line after $i^{\it th}$ scattering, we have,
 
\begin{eqnarray}
 \mbx\!\!\!\!\!\!\!\!\!\!\!\!\!\!
 &&q_{i+1}^2 - M^2  \nn \\
 \mbx\!\!\!\!\!\!\!\!\!\!\!\!\!\! 
 &=& (q+k_i)^2 - M^2  \nn \\ 
 \mbx\!\!\!\!\!\!\!\!\!\!\!\!\!\!
 &=& 2 p^{+}q^{-} \left[  \bar{x}_{i} - x_{B}  - { \bar{x}_D    \mbx}_{i}  + x_{B} \bar{y}_{i}\right]
 \end{eqnarray}
 where we have introduced some momentum fraction variables and  defined a few new variables. These have been defined purely for convenience,

\bea
\bar{x}_{i} &=& \sum_{j=0}^{i} x_{j}, \,\,\,\,\,\, 
x_j = \frac{p_j^+}{p^+}, \\ 
x_B &=&  \frac{Q^2}{2p^+q^-}, \\
 {\bar{x}_{D} \mbx}_i &=&  \frac{(k^{i}_{\perp})^2}{2p^+q^-} = \sum_{j=1}^i {{x}_{D} \mbx}_i, \\ 
{x_{D}}_{i} &=& \frac{ { p^{i}_{\perp} }^{2} + 2 p^{i}_{\perp} \x \sum\limits_{j=0}^{i-1} p^{j}_{\perp} }{ 2 p^{+} q^{-}} , \\
\bar{y}_{i} &=& \sum_{j=1}^{i} y_{j} , \,\,\,\,\,\, 
y_{j} = \frac{p_{j}^{-}}{q^{-}} .
 \end{eqnarray}
 
For all cases where the mass $M$ scales with a higher power of $\lambda$ than $\lambda Q$, the correction to the case of massless quark traversing an extended nuclear medium is suppressed by a factor of $\lambda^{2}$. We seek only the largest corrections $\propto \lambda^{0}$ compared to the propagation of a light quark in an extended medium. Leading corrections to the propagation of a heavy-quark due to multiple scattering in the medium occur for the case when the mass $M \sim Q$ or $M \sim q^{-} \sim \sqrt{\lambda} Q$, {\it i.e.} the mass is of the order of the largest momentum component. Note: this is not the non-relativistic limit where
$M \gg p $ and $p\sim Q$. We refer to this regime where $M \sim p $ as the intermediate momentum region. If $Q$ always refers 
to the hardest scale in the problem, then the high momentum regime is when $M \sim \lambda Q$, and the low momentum regime (equivalent to the non-relativistic regime) is where $M \sim Q/\lambda$. Physically speaking, the intermediate momentum regime for a $b$-quark corresponds to a total energy $E \sim M$; the high energy regime corresponds to the region where $E \gg M$. It is the intermediate momentum regime where all the somewhat surprising results regarding heavy-quark energy loss have been measured and this is the regime, that we will study in greater detail. 

We recall that the exchanged momenta with the medium have momentum components 
$k \equiv [k^{+},k^{-},\vec{k}_{\perp}] \sim [\lambda^{3/2}, \lambda^{2}, \lambda ]Q$. As such, these are somewhat removed from the scale of the mass of the heavy quark. As a result, we separate the terms containing the mass of the heavy-quark from the remaining terms and re-write the entire set of propagator denominators (both cut and uncut lines) as,

%where we can safely neglect ${\bar{x}_{D} \mbx}_i$,   $\sim {\cal O}(\lambda^2)$, in comparison to both $\tau x_B$ and $\bar{\Delta}_i$, $\sim {\cal O}(\lambda^0)$, as well as $ {\bar x}_i$, $\sim {\cal O}(\lambda^1)$.
%With these simplifications on the denominators from all the internal quark lines together with the on-shell delta function for the final heavy-quark line $l_q$ yields,
 %
 \begin{eqnarray}
 {\cal D}_q  &=& \frac{ 2\pi }{(2p^+q^-)^{n+m+1}} \label{denominators} \\
\mbx \times &\prod_{i=0}^{n-1}& \!\!\!\!\! \left(\frac{1}{  \bar{x}_i - x_{B}(1 + \frac{x_{M}}{x_{0}} ) +  x_{B}\bar{y}_{i} - \bar{x}^{i}_{D}-i\e} \right)  \nn \\
 \ata \delta \left[   \bar{x}_n - x_{B}\left(1 + \frac{x_{M}}{x_{0}} \right) +  x_{B}\bar{y}_{n} - \bar{x}^{n}_{D} \right] \nn \\
\mbx \times &\prod^{m-1}_{j=0}& \!\!\!\! \left( \frac{1}{  \bar{x}'_j - x_{B}\left(1 + \frac{x_{M}}{x'_{0}} \right) +  x_{B}\bar{y}'_{j} - \bar{x}'^{j}_{D} + i\e} \right). \nn
 \end{eqnarray}
 %
%where the factor $\eta$ stands for,
% %
% \begin{eqnarray}
% %
% \eta \!\!&&\!\!= \left( \prod_{i=0}^{n} \frac{1}{1+k_i^-/q^-} \right) 
% %
% \left( \prod_{j=0}^{n'-1} \frac{1}{1+{k'_j}^-/q^-} \right)  \label{C-q}.
% %
% \end{eqnarray}
% %
 In so doing, we have retained the leading corrections in $\lambda$ coming from the longitudinal momentum loss experienced by the heavy quark from exchanged gluons with non-negligible $k_{i}^{-}$. In the equation above $x_{M} = M^{2}/(2 p^{+} q^{-} )$. Note that $x_{M} \sim \lambda $ and can be ignored. 
 
 In the high energy ($Q \rightarrow \infty$) and collinear $(\lambda \rightarrow 0)$ limit, we may approximate \cite{Majumder:2007ne},
 \begin{eqnarray}
 \langle~\bar{\psi}(y)~\hat{O}~\psi(0)~\rangle~\approx~\frac{\gamma^-}{2} ~\langle~\bar{\psi}(y)~\frac{\gamma^+}{2}~\hat{O} ~\psi(0)~\rangle.
 \label{approx_high}
 \end{eqnarray}
 As, $A^{+} \sim \lambda^{2} Q$ and $A_{\perp} \sim \lambda^{3} Q$, in $A^{-} = 0$ gauge~\cite{Majumder:2009ge}, we approximate,
 \begin{eqnarray}
 \gamma \cdot A(y) \approx \gamma^- A^+(y). \label{A+}
 \end{eqnarray}
Though we have retained, in Eq.~\eqref{denominators}, the order $\lambda^2$ corrections to the propagators, nonetheless we neglect contributions to the vertices from $A_{\perp}$ in Eq.~\eqref{A+}. 
%The first non-zero contribution will be of the from $\gamma_{\perp} \cdot {q_{i}}_{\perp} \gamma_{\perp} \cdot A_{\perp}$ which is obviously of order $\lambda^{2}$ suppressed compared to leading term. 
However, retained them in the denominators as the order $\lambda^{2}$ terms are the leading terms in the denominators of the propagators. With all these simplifications, structure of the numerator will be as follows, 
 \begin{widetext} 
 \begin{eqnarray}
 \!\!\! && \langle A | \bar{\psi}(y_0) \gamma^\mu \left(\prod_{i=1}^{n} \gamma\cdot (q_i+M)~\gamma\cdot A^{a_i}(y_i) \right) \gamma \cdot l_q \left(\prod_{j=n'}^{1}
 \gamma \cdot  A^{a'_j} (y'_j)~\gamma\cdot (q'_j+M)\right) \gamma^\nu \psi(0) |A\rangle \nonumber\\
 \!\!\! &=& \langle A | \bar{\psi}(y_0) \frac{\gamma^+}{2} \left(\prod_{i=1}^{n} A^{+ a_i}(y_i)\right) \left(\prod_{j=n'}^{1} A^{+ a'_j}(y'_j)\right) \psi(0) |A\rangle \nonumber \\
\ata {\rm Tr} \left[\frac{\gamma^-}{2} \gamma^\mu \left(\prod_{i=1}^{n} \gamma\cdot (q_i+M)~ \gamma^{-}\right) \gamma
 \cdot l_q \left(\prod_{j=n'}^{1} \gamma^{-} ~\gamma\cdot (q'_j+M)\right) \gamma^\nu \right]. \ \ \ \ \ \
 \end{eqnarray}
\end{widetext}

Following $q_{i+1} = q + k_i = q + \sum_{j=0}^i p_i$, the trace now becomes,
 \begin{eqnarray}
 {\cal T} &=& {\rm Tr}\left[\frac{\gamma^-}{2} \gamma^\mu \left(\prod_{i=1}^{n}
 \gamma \cdot (q + k_{i-1}) \gamma^{-} \right)  \right. \label{trace-simplified} \\
 \ata  \left. \gamma \cdot (q + k_n) \left(\prod_{j=m}^{1}
 \gamma^{-} \gamma \cdot (q+k_{j-1}')\right) \gamma^\nu \right], \nn
 \end{eqnarray}
where $\gamma \cdot (q+k_i)  = \gamma^+(q^- + k_i^-)  + \gamma^-(q^+ + k_i^+) - \vec{\gamma}_\perp \cdot \vec{k}_{i}^{\perp}$.
As $\{\gamma^-,\gamma^-\}=0$ the term in $\gamma \cdot (q+k_i)$ containing a $\gamma^-$ vanishes in the trace. 
 
We now obtain,
\bea
{\cal T} &=& \left( g^{\mu}_{~\perp}g^{\nu}_{~\perp} - \frac{M^2}{(q^{-})^2}~g^{\mu}_{~+}g^{\nu}_{~+}  \right) \left( 2q^- \right)^{n+m+1} .  \\
\nn
\eea

In the equation above, we have ignored the suppressed factors of $x_{M}$ as well as those of $p_{0}^{-}$.
While such terms have been dropped from the numerator, they will remain in the denominators  and in the overall $\kd$-function 
until the contour integrations are carried out and the 
denominators expanded in $\lambda$. This is done, to clearly demonstrate that these terms are sub-leading in the determination of the pole structure and in the ensuing expansion. 
One now obtains the contribution to the hadronic tensor from the term with $m$ scatterings in the amplitude and $n$ scatterings in the complex conjugate as (with both leading projections),  
\begin{widetext}
\begin{eqnarray}
W_{mn}^{A\mu\nu} &=& \sum_q~Q_{\cal Q}^2~g^{n+m}~\frac{1}{N_c}~{\rm Tr} \left[ \left( \prod_{i=1}^{n} T^{a_i} \right) \left( \prod_{j=m}^{1} T^{a'_j} \right) \right]
~\int \frac{d^3l_q}{(2\pi)^3}  (2\pi)^3 \delta^3(\vec{l}_q-\vec{q}-\vec{k}_n) \nonumber \\
&\times& \left(\prod_{i=0}^{n} \int dy_i^- \int d^3y_i \right) \left(\prod_{j=1}^{m} \int d{y'}_j^- \int d^3y'_j\right)
\left(\prod_{i=0}^{n} \int \frac{dx_i}{2\pi} \int \frac{d^3p_i}{(2\pi)^3}\right)
\left(\prod_{j=0}^{m-1}  \int \frac{{dx'}_j}{2\pi} \int \frac{d^3p'_j}{(2\pi)^3} \right) \nonumber \\
&\times& \left(\prod_{i = 0}^{n} e^{-ix_i p^+ (y_i^--{y'}_{m}^-)} e^{-i \vec{p}_i \cdot
(\vec{y}_i - \vec{y}'_{m})} \right) \left(\prod_{j=0}^{m-1} e^{i{x'}_j p^+ ({y'}_j^--{y'}_{m}^-)}
e^{i \vec{p}'_j \cdot (\vec{y}'_j - \vec{y}'_{m})} \right) \nonumber \\
%
%\left( 1 - \frac{x_{M}}{x_{0}} \right)^{n+m+1} \nonumber \\
%
&\times&   (2\pi) \delta(- x_B \tau_{M} + \bar{x}_n - \bar{\Delta}_{n} )\left(\prod_{i=0}^{n-1} \frac{1}{-x_B\tau_{M} + \bar{x}_i - \bar{\Delta}_i - i\epsilon}\right)
\left(\prod_{j=0}^{m-1} \frac{1}{- x_B\tau_{M} + \bar{x'}_j - \bar{\Delta}'_j + i\epsilon}\right)  \nonumber\\
&\times& \left( -~g^{\mu\perp}g^{\nu\perp} + \frac{M^2}{(q^-)^2}~g^{\mu -}g^{\nu -}  \right) \langle A | \bar{\psi}(y_0) \frac{\gamma^+}{2}
\left(\prod_{i=1}^{n} A^{+ a_i}(y_i)\right) \left(\prod_{j=m}^{1} A^{+ a'_j}(y'_j)\right) \psi(0) |A\rangle,
\end{eqnarray}
\end{widetext}
where using the delta function $(2\pi) \delta(l_{q}^+ - q^+ - k_n^+)$ one can now able to  perform the integration over ${l_{q}^+}$.
Integration variables have also been changed for convenience, $p_i^+ \to x_i = p_i^+/p^+$, $p_j^{+'} \to x_j' = p_j^{+'}/p^+$.
In the equation above,
\bea
\tau_{M} &=& 1 + x_{M}/x_{0} \simeq 1,  \nn \\ 
\bar{\Delta}_{i} &=& \sum_{j=1}^{i}  \Delta_{j} = { \bar{x}_{D} \mbx}_{i} - x_{B} \bar{y}_{i},    \nn \\
\bar{\Delta}'_{i} &=& \sum_{j=1}^{i}  \Delta'_{j} = { \bar{x}'_{D} \mbx}_{i} - x_{B} \bar{y}'_{i} .
\eea
While the factor $\tau_{M}$ will eventually be set to unity, we retain it in the next few expressions.
We have introduced additional notations for convenience:
$\vec{p} = (p^-, \vec{p}_\perp)$ and $\vec{y} = (y^+, \vec{y}_\perp)$, with $\vec{p}\cdot \vec{y} = p^-y^+ - \vec{p}_\perp \cdot \vec{y}_\perp$.

The end delta function which constrains cut line to be on shell is needed to integrate over $x_n$,
\begin{eqnarray}
\bar{x}_n =   \tau_{M} x_B  +\bar{\Delta}_n ~.
\end{eqnarray}
{\it i.e.},
\begin{eqnarray}
x_n &=& -\bar{x}_{n-1} + \tau_{M} x_B + \bar{\Delta}_n \nn \\
&=& -\sum_{i=1}^{n-1} x_i + \tau_{M} x_B + \bar{\Delta}_n ~.
\end{eqnarray}
Now the (+)-component of the phase factor is as follows,
\begin{eqnarray}
\Gamma^+ &=& e^{-i(\tau_{M} x_B +\bar{\Delta}_n)p^+y_n^-} \left(\prod_{i = 0}^{n-1} e^{-ix_i p^+ (y_i^--y_n^-)}\right) \nn \\
\ata e^{i(\tau_{M} x_B +\bar{\Delta}'_m)p^+{y'}_{n'}^-} \left(\prod_{j=0}^{n'-1} e^{i{x'}_j p^+ ({y'}_j^--{y'}_{m}^-)} \right) \nonumber \\ 
&=& \Gamma_n^+ \Gamma_m^+ .
\end{eqnarray}
Two phase factors $\Gamma_n^+$ and $\Gamma_m^+$, respectively, are related to  $x_i$ integration and
the $x'_j$ integration. The rest of the integration may now be performed 
(over the two momentum fractions $x_i$ and $x'_j$).
Starting from the propagators that are attached to the cut line one proceeds to the initial hard scattering electromagnetic vertex.
The integrations over $x_i$ in the complex conjugate ($x_i$'s) will be demonstrated in detail now;
the integrations over the momentum fractions ($x'_j$'s) in the amplitude are similar.

We close the contour of $x_{n-1}$ with a anti-clockwise semi-circle in the upper half of  the complex plane for the integration over the momentum fraction $x_{n-1}$,
\begin{eqnarray}
\int&& \!\!\!\!\!\!\!\frac{dx_{n-1}}{2\pi} \frac{e^{-ix_{n-1} p^+ (y_{n-1}^{-} -y_n^-)}}{-\tau_{M} x_B + x_{n-1}
+ \bar{x}_{n-2}  - \bar{\Delta}_{n-1}- i\epsilon} \nonumber  \\ 
&=& i \theta(y_n^--y_{n-1}^-) \\ 
\ata e^{-i (-\bar{x}_{n-2} + \tau_{M} x_B + \bar{\Delta}_{n-1})p^+ (y_{n-1}^- - y_n^-)}. \nonumber
\end{eqnarray}
The $\theta$-function mimics the fact that the heavy quark  is traveling from $y_{n-1}^-$ to $y_n^-$.
The phase factor together with the above results from the contour integration now becomes,
\begin{eqnarray}
\Gamma_n^+ &\to& e^{-i\bar{\Delta}_n p^+y_n^{-}} e^{-i(\tau_{M} x_B +\bar{\Delta}_{n-1}) p^+ y_{n-1}^-}  \nn \\ 
\ata \left(\prod_{i = 0}^{n-2} e^{-ix_i p^+ (y_i^--y_{n-1}^-)}\right).
\end{eqnarray}
Rest of the integrations over the longitudinal momentum fractions $x_i$'s can be performed in an identical way. Finally, we obtain
\begin{eqnarray}
\Gamma_n^+ &\to& e^{-i \tau_{M} x_B p^+ y_0^-}  \left(\prod_{i=1}^n e^{-i { \Delta}_i p^+ y_i^-} \right) \nn \\
\ata i^n \left(\prod_{i=1}^n \theta(y_i^--y_{i-1}^-) \right).
\end{eqnarray}
The momentum fraction ($x'_j$'s) integrations in the amplitude are identical. Except a factor of $(-i)$ apear in amplitude instead of $i$. This comes from the contour integration with a clockwise semi-circle in the lower half of the plane (for the amplitude) instead of anti clockwise semi-circle in the upper half of the plane.

In all remaining expressions, we will ignore the factors of $\tau_{M} \simeq 1$. After performing all integrations over the internal quark momentum components and taking the Dirac trace of the factors in the numerator, the expression for the hadronic tensor is now much simplified. Combining both the leading projections, we obtain the contribution to the hadronic tensor from the term with 
$m$ scatterings in the amplitude as well as $n$ scatterings in the complex conjugate as follows,
\begin{widetext}
\begin{eqnarray}
&& W_{mn}^{A\mu\nu} \nn \\
&=& \sum_q ~Q_{\cal Q}^2~ g^{n+m}~ \frac{1}{N_c}~ {\rm Tr}\left[\left(\prod_{i=1}^{n} T^{a_i}\right) \left(\prod_{j=m}^{1} T^{a'_j}\right) \right] 
\int \frac{d^3l_q}{(2\pi)^3}  (2\pi)^3 \delta^3(\vec{l}_q-\vec{q}-\vec{k}_n ) \nonumber \\
&\times& \left(\prod_{i=0}^{n} \int dy_i^- \int d^3y_i \right) \left(\prod_{j=1}^{m} \int d{y'}_j^- \int d^3y'_j\right) \left(\prod_{i=0}^{n} \int \frac{d^3p_i}{(2\pi)^3}\right)
\left(\prod_{j=0}^{m-1}  \int \frac{d^3p'_j}{(2\pi)^4} \right)
\left(\prod_{i = 0}^{n} e^{-i \vec{p}_i \cdot \vec{y}_i } \right) \left(\prod_{j=1}^{m}  e^{i \vec{p}'_j \cdot \vec{y}'_j} \right)  \nonumber\\
&\times&  e^{-i x_B  p^+ y_0^-}  \left(\prod_{i=1}^n e^{-i {\Delta}_i p^+ y_i^-} \right) \left(\prod_{j=1}^{m} e^{i {\Delta'}_j p^+ {y'}_j^-} \right) i^n (-i)^{m} \left(\prod_{i=1}^n \theta(y_i^--y_{i-1}^-) \right)
\left(\prod_{j=1}^{m} \theta({y'}_j^- - {y'}_{j-1}^-) \right) \nonumber\\
&\times& \left(-g_\perp^{\mu\nu} + \frac{M^2}{(q^-)^2}~g^{\mu -}g^{\nu -}  \right)
\langle A | \bar{\psi}(y_0) \frac{\gamma^+}{2} \left(\prod_{i=1}^{n} A^{+ a_i}(y_i)\right) \left(\prod_{j=m}^{1} A^{+ a'_j}(y'_j)\right) \psi(0) |A\rangle.
\end{eqnarray}
\end{widetext}

From the above expression of $W_{mn}^{A\mu\nu}$ its evident that when $M\sim \sqrt{\lambda}Q$ ($q^{-} \sim \sqrt{\lambda}Q$), the two tensor projections $W_{mn}^{A\perp \perp}$ and $W_{mn}^{A+ +}$ are of same order. In the subsequent section we will expand it 
as a series in the small scattering momenta around the hard part of the above expression. 
% ( $W_{mn}^{A+ +}$ was ${\cal O}(\lambda^2)$ suppressed for light quark, $M \sim \lambda^2 Q$)  Its also rather hilarious to see that that $W_{mn}^{A- -}$, $W_{mn}^{A + -}$ did not survive at this order as evident from the expression. This might be an artifact coming from the high energy approximation taken in 
%Eq.[\ref{approx_high}]. One however could easily recover them, thanks to current conservation relations,  
%i.e. $q^-W^{++}_0+q^+W^{-+}_0=0$, $q^-W^{+-}_0+q^+W^{--}_0=0$, $q^-W^{++}_0+q^+W^{+-}_0=0$ and $q^-W^{-+}_0+q^+W^{--}_0=0$.  

%%%%%%%%%%%%%%%%%%%%%%%%%%%%%%%%%%%%%%%%%%%%%%%%%%%%%%%%%%
%%%%%%%%%%%%%%%%%%%%%%%%%%%%%%%%%%%%%%%%%%%%%%%%%%%%%%%%%%
%%%%%%%%%%%%%%%%%%%%%%%%%%%%%%%%%%%%%%%%%%%%%%%%%%%%%%%%%%

\section{Factorization, Gradient expansion and resummation}

%%%%%%%%%%%%%%%%%%%%%%%%%%%%%%%%%%%%%%%%%%%%%%%%%%%%%%%%%%
%%%%%%%%%%%%%%%%%%%%%%%%%%%%%%%%%%%%%%%%%%%%%%%%%%%%%%%%%%
%%%%%%%%%%%%%%%%%%%%%%%%%%%%%%%%%%%%%%%%%%%%%%%%%%%%%%%%%%

In the preceding section, leading hadronic tensor components was elucidated, however, with no approximations regarding the internal structure of the nucleus. In this section, we will make a few pragmatic assumptions on the nuclear states to simplify the structure of the hadronic tensor components. We will consider the special case when $n=m$, {\it i.e.}, the 
number of scatterings in the amplitude is equal to the number of scattering in the complex conjugate amplitude. The cases when they are unequal are generally associated with higher order twist matrix elements. This could also constitutes the unitarity corrections to results where the propagating heavy quark encounters scatterings $\min(n,n')$ times \cite{Majumder:2007hx}.

In this study the nucleus is assumed as a very weakly interacting homogeneous gas of partons. In the high energy limit, such approximation is well justified, where partons supposed 
to travel in straight trajectories and due to time dilation they are almost independent of each other over the time window of the interactions of the leading heavy quark.
Employing this approximation we may now at a position to factorise the expectations of field operators in the {\it nuclear states} into the products of expectations in the {\it nucleon states}. Nucleon being a color neutral object, any combination of parton field strength insertions must have to be to a color singlet combination.
Accordingly, the first non-trivial leading contribution comes from the terms where $2n$ gluons are combined, into $n$ number of color singlet pairs, in diferent nucleon states with one gluon of the pair is from the amplitude and one is from the complex conjugate of the amplitude.

The expectation average of the color fields in the nuclear state is therefore factorised using the time ordered products of $\theta$ functions as,

\begin{eqnarray}
 \langle A | \bar{\psi}(y_0) && \!\!\!\!\!\!\!\!\!\!\! \frac{\gamma^+}{2} \prod_{i=1}^{n} A^{+ a_i}(y_i) \prod_{j=n}^{1} A^{+ a'_j}(y'_j) \psi(0) |A\rangle  \nn \\
&=& C_{p_0, p_1 \cdots p_n} \langle p | \bar{\psi}(y_0) \frac{\gamma^+}{2} \psi(0) |p\rangle \nn \\
\ata \left(\prod_{i=1}^{n} \langle p| A^{+ a_i}(y_i) A^{+ a'_i}(y'_i)|p\rangle \right).
\end{eqnarray}

Under the strict constraints coming from the string of $\theta$-functions the $y_i$ integrations are carried out over the whole nuclear volume. The probability to find $n+1$ nucleons in the proximity of the positions $y_0, y_1 \cdots y_n$ is normalized by the the factor $C_{p_0, p_1 \cdots p_n}$. Within the approximations mentioned above this factor may be estimated as
\begin{eqnarray}
C_{p_0, p_1 \cdots p_n} = A C_p^A \left(\frac{\rho}{2p^+}\right)^n ,
\end{eqnarray}
here $\rho$ being the parton density inside the large nucleus, and the factor $1/(2p^+)$ is require for the normalization of partonic state. For a nucleus with somewhat inhomogeneous density the normalization coefficient will possess explicite spatial dependence. One may now make an average over the colors of the gauge fields,
\begin{eqnarray}
\langle A^a(y) A^b(0)\rangle = \frac{\delta_{ab}}{N_c^2 - 1} \langle A(y) A(0) \rangle.
\end{eqnarray}
The average over the colors of quark field brings the factor $1/N_c$. Therefore, the overall color factors is,
\begin{eqnarray}
&&\frac{1}{N_c} \frac{1}{(N_c^2 -1)^n} {\rm Tr}\left[\left(\prod_{i=1}^{n} T^{a_i}\right) \left(\prod_{j=n}^{1} T^{a_i}\right) \right]  \nn \\
&&= \left(\frac{C_F}{N_c^2-1}\right)^n.
\end{eqnarray}
Leading components of the hadronic tensor will now become,
\begin{widetext}
\begin{eqnarray}
W_{nn}^{A\mu\nu}  &=& \sum_q ~ Q_q^2 ~ \left(-g_\perp^{\mu\nu} + \frac{M^2}{(q^-)^2}~g^{\mu -}g^{\nu -}  \right)~ 
 ~ A ~ C_p^A \left(\frac{\rho}{2p^+}\right)^n g^{2n} \left(\frac{C_F}{N_c^2 - 1} \right)^n \nonumber \\
&\times& \int \frac{d^3l_q}{(2\pi)^3}  (2\pi)^3 \delta^3(\vec{l}_q-\vec{q}-\vec{k}_n ) \int dy_0^- \int d^3y_0 \int \frac{d^3p_0}{(2\pi)^3} e^{-i \vec{p}_0 \cdot \vec{y}_0} e^{-i \tau x_B p^+ y_0^-}  \langle p | \bar{\psi}(y_0) \frac{\gamma^+}{2} \psi(0) |p\rangle \nonumber \\
&\times&  \left(\prod_{i=1}^{n} \int dy_i^- \int d{y'}_i^- \theta(y_i^--y_{i-1}^-) \theta({y'}_i^- - {y'}_{i-1}^-) \right)\left(\prod_{i=1}^{n} \int d^3y_i \int d^3y'_i \int \frac{d^3p_i}{(2\pi)^3} \int \frac{d^3p'_i}{(2\pi)^3} \right. \nonumber \\ 
&\times&  \left. e^{-i \vec{p}_i \cdot \vec{y}_i }  e^{i \vec{p}'_i \cdot \vec{y}'_i} e^{-i {\Delta}_i p^+ y_i^-}  e^{i {\Delta}'_i p^+ {y'}_i^-} \langle p| A^{+}(y_i) A^{+}(y'_i)|p\rangle \right)~ ,
\end{eqnarray}
\end{widetext}

where the integrating variable $dp'_0$ have been changed to $dp'_n$. Using the homogeneity approximation the expression should be further simplified by following transformation of variables $(y_i, y'_i) \to (Y_i, \delta y_i)$,
\begin{eqnarray}
Y_i = (y_i + y'_i)/2, \ \ \ \delta y_i = y_i - y'_i .
\end{eqnarray}
Translational invariance over an extended spatial dimension for a large nucleus,  allow us to express the expectation values of gluon operators as,
\begin{eqnarray}
\langle p| A^{+}(y_i) A^{+}(y'_i)|p\rangle  &\simeq& \langle p|A^+(y_i-y'_i) A^+(0) |p\rangle   \nn \\
&\simeq& \langle p|A^+(\delta y_i) A^+(0) |p\rangle . 
\end{eqnarray}
It is now possible to perform the integration over the phase factor, which now depends only on the average values $\vec{Y}_i$, 

%\begin{eqnarray}
%&&\int d^3y_i \int d^3y_i{\hspace{0.1mm}'  e^{-i \vec{p}_i \cdot %\vec{y}_i }  e^{i \vec{p}_i{\hspace{0.1mm}'} \cdot \vec{y}_i{\hspace{0.1mm}'}  \nonumber \\
%&&= \int d^3Y_i \int d^3 \delta y_i e^{-i (\vec{p}_i-\vec{p}_i{\hspace{0.1mm}'}) \cdot \vec{Y}_i } e^{-i (\vec{p} + \vec{p}_i{\hspace{0.1mm}'}) \cdot \delta \vec{ y}_i/2}  \nonumber \\
%&&= (2\pi)^3 \delta^3(\vec{p}_i-\vec{p}_i{\hspace{0.1mm}'}) \int d^3 \delta y_i e^{-i \vec{p} \cdot \delta \vec{y}_i}. \nonumber 
%\end{eqnarray}

\begin{eqnarray}
&&\int d^3y_i \int    d^3y_i{\hspace{0.1mm}'}  e^{-i \vec{p}_i \cdot \vec{y}_i }  e^{i \vec{p}_i{\hspace{0.1mm}'} \cdot \vec{y}_i{\hspace{0.1mm}'}} \nonumber \\
&&= \int d^3Y_i \int d^3 \delta y_i e^{-i (\vec{p}_i-\vec{p}_i{\hspace{0.1mm}'}) \cdot \vec{Y}_i } e^{-i (\vec{p} + \vec{p}_i{\hspace{0.1mm}'}) \cdot \delta \vec{ y}_i/2}  \nonumber  \\
&&= (2\pi)^3 \delta^3(\vec{p}_i-\vec{p}_i{\hspace{0.1mm}'}) \int d^3 \delta y_i e^{-i \vec{p} \cdot \delta \vec{y}_i}. \nonumber 
\end{eqnarray}

This delta function actually fix the momentum fractions ${x_{D}}_i = {x_{D}'}_i$.
Since both $\delta y_i^-$ and $\delta y_{i-1}^-$ belongs to the nucleon size (smaller compared to the size of the large nucleus $\sim$ $Y^-$), we may simplify product of $\theta$-functions as,
\begin{eqnarray}
\theta(y_i^--y_{i-1}^-) \theta({y'}_i^- - {y'}_{i-1}^-) = \theta(Y_i^- - Y_{i-1}^-).
\end{eqnarray}
The time-ordered product of $\theta$-functions is now as follow,
\begin{eqnarray}
\prod_{i=1}^{n} \int_0^{L^-} \!\!\!\!\!dY_i^- \theta(Y_i^- - Y_{i-1}^-)
&=& \frac{ \prod\limits_{i=1}^{n} \int\limits_0^{L^-} dY_i^- }{n!},
\end{eqnarray}
where extent of the nuclear size is expressed by $L^-$. These terms, on integration, yield a factor of ${ (L^{-} )}^{n}$ and cause the overall length enhancement of the process. 
Finally we obtain the leading components of the differential hadronic tensor for $n$ scatterings both in the amplitude and complex conjugate as,

\begin{widetext}

\begin{eqnarray}
\frac{d W_{nn}^{A\mu \nu}}{{d^3l_q} }  &=&  \sum_q Q_q^2 \left(-g_\perp^{\mu\nu} + \frac{M^2}{(q^-)^2}~g^{\mu -}g^{\nu -}  \right)
 A C_p^A \int dy_0^- e^{-i x_B p^+ y_0^-} \langle p | \bar{\psi}(y_0) \frac{\gamma^+}{2} \psi(0) |p\rangle \nonumber \\
&\times& \frac{1}{n!}~\prod_{i=1}^{n}~\left( \int_0^{L^-} dY_i^- \int d\delta y_i^- \int d^3 \delta y_i \int \frac{d^3p_i}{(2\pi)^3} \frac{\rho}{2p^+} g^{2} \frac{C_F}{N_c^2 - 1}
e^{-i \vec{p}_i \cdot \delta \vec{y}_i} \langle p| A^{+}(\delta y_i) A^{+}(0) |p\rangle \right) \nonumber \\
&\times&  \left(\prod_{i=1}^{n}  e^{-i {\Delta}_i p^+ \delta y_i^-}  \right) \delta^3 \left( \vec{l}_q-\vec{q}  - \sum_{i=1}^n \vec{p}_i \right)  \nn \\
&=& \left(W_{0\perp}^{A\mu \nu} + W_{0 L}^{A\mu \nu} \right)  \phi_{n} . \label{n-scat-general-form}
\end{eqnarray}
In the equation above, $W_{0\perp}^{A\mu \nu}$ and $W_{0 L}^{A\mu \nu} $ represent the leading order hadronic tensors 
without any rescattering of the produced heavy quarks. They are given as, 
\bea
W_{0\perp}^{A\mu \nu} &=& \left(-g^{\mu\perp}g^{\nu\perp} \right) C_p^A\sum_q Q_q^2  \int dy_0^- e^{-i x_B p^+ y_0^-} \langle p | \bar{\psi}(y_0) \frac{\gamma^+}{2} \psi(0) |p\rangle,
\eea
and for the leading light-cone projection as, 
\bea
W_{0 L }^{A\mu \nu} &=& \left( g^{\mu -}g^{\nu -}\frac{M^2}{(q^-)^2} \right) C_p^A\sum_q  Q_q^2 ~  \int dy_0^- e^{-i x_B p^+ y_0^-} \langle p | \bar{\psi}(y_0) \frac{\gamma^+}{2} \psi(0) |p\rangle.
\eea
The factor $\phi_{n}$ in Eq.~\eqref{n-scat-general-form} represents the piece from $n$-scattering on the outgoing heavy quark in the final state. This contains both the ``hard-part'' which contains 
factors of the momentum of the heavy-quark, as well as the ``soft-part'' which contains phase factors and nucleon matrix elements.

We are now in a position to go for the resummation over multiple scatterings. To do so we will adopt further simplifications coming from the collinear/eikonal approximations, where the exchanged momenta are small compared to momenta of the leading heavy quark.
Assuming that the hadronic tensor is analytic around $\vec{p}_i = 0$ we will Taylor expand it around the soft exchanged momenta,
\begin{eqnarray}
H (q^-,p^+,p_0^-,p_i^\alpha) = \prod_{i=1}^{n} \left[\left(H\right)_{\vec{p}_1 \cdots \vec{p}_n = 0} + p_i^\alpha \left(\frac{\partial}{\partial p_i^\alpha}H\right)_{\vec{p}_1 \cdots \vec{p}_n = 0} + \frac{1}{2} p_i^\alpha p_i^\beta \left(\frac{\partial}{\partial p_i^\alpha} \frac{\partial}{\partial p_i^\beta}H\right)_{\vec{p}_1 \cdots \vec{p}_n = 0} + \cdots \right] .
\end{eqnarray}
In the above expansion, the $\alpha, \beta$ represents both ``$-$'' and ``$\perp$''. Terms up to the second order have been retained for simplicity. The higher derivative terms in the expansion actually correspond to higher order moments of the exchanged momentum distribution. 

In the above expansion, the first terms (the term without any derivative) generally are gauge corrections for the diagrams with lower number of scatterings.  In a manifestly gauge invariant expression its important to include those terms also. 

The exchanged momentum may now be transformed into the appropriate derivatives over position,
\begin{eqnarray}
 e^{-i\vec{p}_i \cdot \delta\vec{y}_i} \langle p| A^+(\delta \vec{y}_i) A^+(0)|p \rangle p_i^\alpha \frac{\partial}{\partial p_i^\alpha} &=&   e^{-i\vec{p}_i \cdot \delta\vec{y}_i} (-i) \langle p| \partial^\alpha A^+(\delta \vec{y}_i) A^+(0)|p \rangle \frac{\partial}{\partial p_i^\alpha} , \nonumber\\
 e^{-i\vec{p}_i \cdot \delta\vec{y}_i} \langle p| A^+(\delta \vec{y}_i) A^+(0)|p \rangle p_i^\alpha p_i^\beta \frac{\partial}{\partial p_i^\alpha} \frac{\partial}{\partial p_i^\beta} &=& e^{-i\vec{p}_i \cdot \delta\vec{y}_i} \langle p| \partial^\alpha A^+(\delta \vec{y}_i) \partial^\beta A^+(0)|p \rangle \frac{\partial}{\partial p_i^\alpha} \frac{\partial}{\partial p_i^\beta} .
\label{Hexp}
\end{eqnarray}
It is now straightforward to perform the integrations over $\vec{p}_i$ and $\delta\vec{y}_i$. We may ignore the terms coming from the derivatives of the phase factor $e^{-i {x_{D}}_i p^+ \delta y_i^-}$.  It will essentially results in spatial moments of the two gluon field products such as $\langle A^+(\delta y^-, \delta \vec{y}_{\perp}) \delta y^- A^+(0) \rangle$. However we will keep the leading term arising from the delta functions when evaluating the momentum derivatives on the hard part.
The transverse projection of the differential hadronic tensor now reads as,
\begin{eqnarray}
\frac{d W_{nn \perp}^{A\mu \nu}}{{d^3l_q} }  \!\!&&\!\! =  W_{0\perp}^{A\mu \nu }
\frac{1}{n!} \left(\prod_{i=1}^{n} \int_0^{L^-} \!\!\!\!\!\!\!dY_i^-  \left[ - {\cal D}_{L1} \frac{\partial}{\partial p_i^-} + \frac{1}{2} {\cal D}_{L2} \frac{\partial^2}{\partial^2 p_i^-} + \frac{1}{2} {\cal D}_{D2} {\nabla_{p_{i \perp}}^2}  \right] \right)  \delta^3 \left( \vec{l}_q - \vec{q} - \sum_{i=1}^n \vec{p}_i \right)_{\vec{p}_1 \cdots \vec{p}_n = 0} \!\!\!\!\!\!\!\!\!\!\!\!\!\!. 
\end{eqnarray}
The hard sector transport coefficients ${\cal D}_{L1}$, ${\cal D}_{L2}$ and ${\cal D}_{T2}$ are defined as,

\begin{eqnarray}
{\cal D}_{L1} \!\!&&\!\! = g^2 \frac{C_F}{N_c^2 - 1} \int dy^- \frac{\rho}{2p^+} \langle p| i\partial^- A^+(y^-) A^+(0)|p \rangle\left(\prod_{i=1}^{n}  e^{-i {\bar \Delta}_i p^+ \delta y_i^-}  \right) , \nonumber\\
{\cal D}_{L2} \!\!&&\!\! = g^2 \frac{C_F}{N_c^2 - 1} \int dy^- \frac{\rho}{2p^+} \langle p| \partial^- A^+(y^-) \partial^- A^+(0)|p \rangle\left(\prod_{i=1}^{n}  e^{-i {\bar \Delta}_i p^+ \delta y_i^-}  \right),
\nonumber\\
{\cal D}_{T2} \!\!&&\!\! = g^2 \frac{C_F}{N_c^2 - 1} \int dy^- \frac{\rho}{2p^+} \langle p| \partial_\perp A^+(y^-) \partial_\perp A^+(0)|p \rangle\left(\prod_{i=1}^{n}  e^{-i {\bar \Delta}_i p^+ \delta y_i^-}  \right). \ \ \ \ \
\end{eqnarray}

These three coefficients
${\cal D}_{L1}$, ${\cal D}_{L2}$ and ${\cal D}_{D2}$ are connected to longitudinal energy loss rate $\hat{e}$, longitudinal momenta diffusion rate $\hat{e}_2$ and transverse momenta diffusion rate $\hat{q}$. Its is also worth mentioning that, while for light quark ${\bar \Delta} \sim {\bar x_D} \sim \lambda^2$, for the `semi hard' heavy quark ${\bar \Delta} \sim {\bar x_D}-x_{\small B}{\bar y_i}  \sim \lambda^{\frac{3}{2}}$. As such, the hard sector transport coefficients of a heavy-quark sample somewhat higher values of momentum fraction $x$ than light quark transport coefficients. This supports the notion that heavy quark and light quark transport coefficients need not yield the same numerical value. The above definitions of all the hard sector transport coefficients are not truly gauge-invariant. The manifestly gauge invariant hard sector transport coefficients can only be realised with the incorporation of higher order much softer terms where $k_{\perp} \ll \lambda Q$. The summation over such ultra soft gluon incorporation eventually leads to the emergence of Wilson links between the the gluon field operators.  This will renders the operator product gauge invariant. 

We have now resummed over an arbitrary number of multiple scatterings,
\begin{eqnarray}
\frac{d W^{A\mu \nu}_{\perp}}{{d^3l_q} } = \sum_{n=0}^{\infty} \frac{d W_{n \perp}^{A\mu \nu}}{{d^3l_q} }  =  W_0^{A\mu \nu }
\phi (L^-, l_q^-, \vec{l}_{q\perp}),  \ \ \ \ \ \
\end{eqnarray}
where the final state quark distribution function is defined as $\phi(L^-, l_q^-, \vec{l}_{q\perp})$,
\begin{eqnarray}
\phi \!\!&&\!\! = \sum_{n=0}^{\infty} \frac{1}{n!} \left(\prod_{i=1}^{n} \int_0^{L^-} dY_i^- \left[ - {\cal D}_{L1}\frac{\partial}{\partial p_i^-} + \frac{1}{2} {\cal D}_{L2} \frac{\partial^2}{\partial^2 p_i^-}  + \frac{1}{2} {\cal D}_{T2} {\nabla_{p_{i \perp}}^2}  \right] \right)~\delta(l_q^- - q^- - p_0^-)~\delta^2(\vec{l}_{q\perp}). \ \ \ \ \
\end{eqnarray}
The heavy quark momentum distribution function, after the full multi-scattering resummation reads,
\begin{eqnarray}
\phi(L^-, l_q^-, \vec{l}_{q\perp}) \!\!&&\!\! = \exp\left( L^- \left[ {\cal D}_{L1}\frac{\partial}{\partial l_q^-} + \frac{1}{2} {\cal D}_{L2} \frac{\partial^2}{\partial^2 l_q^-} + \frac{1}{2} {\cal D}_{T2}{\nabla_{l_{q\perp}}^2}  \right] \right)~\delta(l_q^- - q^- - p_0^-)~\delta^2(\vec{l}_{q\perp}),
\end{eqnarray}
where the derivatives over $p_i$ have been transformed to the derivatives over $l_q$.

The momentum distribution function $\phi(L^-, l_q^-, \vec{l}_{q\perp})$ for the final outgoing heavy quark is the unique solution of the following diffusion equation,
\begin{eqnarray}
\frac{\partial \phi}{\partial L^-} = \left[ {\cal D}_{L1}\frac{\partial}{\partial l_q^-} + \frac{1}{2} {\cal D}_{L2} \frac{\partial^2}{\partial^2 l_q^-} + \frac{1}{2} {\cal D}_{T2}{\nabla_{l_{q\perp}}^2}  \right] \phi(L^-, l_q^-, \vec{l}_{q\perp}).
\end{eqnarray}
Above mentioned differential equation describes the time evolution of the  momentum distribution profile of propagating heavy quark which suffers multiple soft scatterings in the passage of its transport through a nuclear matter.
The three terms in the above diffusion equation represent the contributions from longitudinal momentum change and longitudinal momentum diffusion, and the  transverse momentum diffusion.
The delta function initial condition: $\phi(L^- = 0, l_q^-, \vec{l}_{q\perp}) =  \delta(l_q^- - q^-) \delta^2(\vec{l}_{q\perp})$, provied the following solution for the distribution fuction $\phi$,
\begin{eqnarray}
\phi(L^-, l_q^-, \vec{l}_{q\perp}) = \frac{1}{\sqrt{2\pi {\cal D}_{L2} L^-}}\exp\left[-\frac{{(l_q^- - q^- + {\cal D}_{L1} L^-)^2}}{{2{\cal D}_{L2}L^-}}\right] ~~~ \frac{1}{2\pi {\cal D}_{T2} L^-}\exp\left[\frac{-{{l}_{q\perp}^2 }}{{2{\cal D}_{T2}L^-}}\right] .
\end{eqnarray}
\end{widetext}
One may now identify,
\begin{eqnarray}
\langle l_q^- \rangle  &=& q^- - {\cal D}_{L1} L^-, \\
\langle (l_q^-)^2 \rangle - \langle l_q^- \rangle^2 &=& {\cal D}_{L2} L^-, \nn \\
\langle l_{q\perp}^2 \rangle  &=& 2 {\cal D}_{T2} L^- .
\end{eqnarray}
The coefficients $D_{L1}$, $D_{L2}$ and $D_{T2}$ are related to longitudinal  drag rate $\hat{e}=dE/dt$, longitudinal straggling rate $\hat{e}_2 = d(\Delta E)^2/dt$, and the transverse momentum diffusion rate $\hat{q} = d(\Delta p_T)^2/dt$ as
\begin{eqnarray}
\hat{e} &=& {\cal D}_{L1}, \nn \\
\hat{e}_2 &=& {\cal D}_{L2}/\sqrt{2}, \nn \\
\hat{q} &=& 2\sqrt{2} {\cal D}_{T2}.
\end{eqnarray}

A similar set of arguments may be used to simplify the $(++)$-projection of the hadronic tensor.
At this order of approximation, we obtain the longitudinal projection of the hadronic tensor from an arbitrary number of scatterings as,
\begin{eqnarray}
\frac{d W^{A\mu \nu}_{L}}{{d^3l_q} } = \sum_{n=0}^{\infty} \frac{d W_{n L}^{A\mu \nu}}{{d^3l_q} }  =  W_{0L}^{A\mu \nu }
\phi (L^-, l_q^-, \vec{l}_{q\perp}), 
\end{eqnarray}

\section{Conclusions}
The subject of heavy quark energy loss is not yet a settled issue and requires more detailed analysis \cite{Mustafa:2004dr, Bhattacharyya:2013hga}. In this work, the propagation of a single `semi-hard' heavy quark in the dense nuclear medium have been studied within an extension of higher twist framework that includes multiple scatterings also. In this formalism the higher twist corrections, magnified by the large extent of the nucleus, are resummed to obtain the temporal evolution for the momentum distribution of the `semi hard' heavy quark. Both transverse momentum broadening as well as the longitudinal drag and longitudinal momentum diffusion of the heavy quark, have been studies simultaneously within this unified framework of higher twist formalism for multiple scatterings.

We have focussed on the specific case of ``semi-hard'' quarks where the mass and momentum scale as $M, p \sim \sqrt{\lambda} Q$. SCET-Glauber scaling based momentum power counting shows that the longitudinal momentum transfers 
and the transverse momentum transfers have a comparable effect on the off-shellness of the heavy-quark. 
%that leads the energy loss heavy flavor in the intermediate energy scale 
%unlike the light quark where diffusion contribute proactively on jet energy loss. 
This implies that longitudinal transfers, not only lead to the drag and diffusion of non-radiating quarks in a medium, as is the case for light flavors, but will also noticeably affect the radiative loss. 
The calculation of this novel effect, will be carried out in a future effort. In this paper, we focussed on whether the drag and diffusion experienced by heavy-flavors can be cast in the same form as for light flavors; this is indeed the case. 

An evolution equation for the temporal development of the heavy-quark momentum distribution, as guided by the multiple scatterings from the dense medium, have been derived. All three leading transport coefficients involved are connected to the longitudinal drag, longitudinal straggling and transverse momentum diffusion coefficient. The general structure of the transport coefficients for the semi-hard heavy quarks appear to be similar to those for light quarks (or even those for a fast heavy-quark). However, a closer analysis of the ratio $x$ of the ($+$)-component of the momentum of the exchanged gluon with that of the target nucleon, indicates that semi-hard heavy quarks,  scattering off the nucleon, sample a larger value of $x$ than do light quarks or gluons. As such, the values of the transport coefficients for such partons may not be the same as those used for light flavors. 
As a corollary, the energy loss of semi-hard heavy-quarks yield a direct window into the $x$ dependence of jet transport coefficients. The combined effect of all the three hard sector transport coefficients on the gluon bremsstrahlung spectrum off the heavy quark will be explored in a future effort.

\acknowledgements
 The authors would like to thank G.-Y. Qin for helpful discussions.
 This work was supported in part by the National Science Foundation under grant number PHY-1207918. This work is also supported in part by the Director, Office of Energy Research, Office of High Energy and Nuclear Physics, Division of Nuclear Physics, of the U.S. Department of Energy, through the JET topical collaboration.

%\begin{center} 
%NEED TO REMOVE the following part 
%\end{center}  
% 
%
%
%\begin{eqnarray}
%e_{\perp\perp} &=& g_{\perp\perp} - \frac{q_\perp q_\perp}{q^2} ,  \nonumber \\
%               &=& g_{\perp\perp}.  \\
%%
%d_{\perp\perp} &=& - g_{\perp\perp} - \frac{p_\perp p_\perp}{(p.q)^2} q^2  + \frac{p_\perp q_\perp +p_\perp q_\perp}{p.q}, \nn \\
%&=& - g_{\perp\perp} + {\cal O}(\lambda^2).  \\
%\end{eqnarray}
%
%
%\begin{eqnarray}
%e_{++} &=& g_{++} - \frac{q_+ q_+}{q^2} ,  \nonumber \\
%               &=&  - \frac{q^+}{q^-}~.  \\
%%
%d_{++} &=& - g_{++} - \frac{p^+ p^+}{(p.q)^2} q^2  + \frac{p_+ q_+ +p_+ q_+}{p.q}, \nn \\
%&=& 4\xi^2~\frac{(p^+)^2}{Q^2}  - 4\xi~\frac{p^+}{q^-}~.  \\
%\end{eqnarray}
%
%
%\begin{eqnarray}
%e_{--} &=& g_{--} - \frac{q^- q^-}{q^2} \nonumber \\
%               &=&  \frac{(q^-)^2}{Q^2}~.  \\
%%
%d_{--} &=& - g_{--} - \frac{p^- p^-}{(p.q)^2} q^2  + \frac{p^- q^- +p^- q^-}{p.q}, \nn \\
%&=& 2\xi~\frac{M^2}{Q^2}~\frac{q^-}{p^+}+4 \xi^2~\frac{M^2}{Q^2 }~\frac{M^2}{(p^+)^2}  ~.  \\
%\end{eqnarray}
%
%
%\begin{eqnarray}
%e_{+-} &=& g_{+-} - \frac{q^+ q^-}{q^2} \nonumber \\
%               &=& 0~.  \\
%%
%d_{+-} &=& - g_{+-} - \frac{p^+ p^-}{(p.q)^2} q^2  + \frac{p^+ q^- +p^- q^+}{p.q}, \nn \\
%&=& 4 \xi^2~\frac{M^2}{Q^2}~.  \\
%\end{eqnarray}
%
%
%
%
%
%

\end{document}